\documentclass[12pt,fleqn]{article}
\usepackage[T1]{fontenc}
\usepackage[ansinew]{inputenc}
\usepackage{lmodern}
\usepackage{graphicx}
\usepackage{epstopdf}
\usepackage{amsmath}
\usepackage{amsthm}
\usepackage{amsfonts}
\usepackage{indentfirst}
\usepackage{amssymb}
\usepackage{setspace}
\usepackage{textcomp}
\usepackage{float}
\usepackage{url}
\usepackage{color}
\renewcommand{\theequation}{\arabic{section}.\arabic{equation}}
\newcommand{\ba}{\begin{array}}
\newcommand{\ea}{\end{array}}

\begin{document}
\newcommand{\be}{\begin{equation}}
\newcommand{\ee}{\end{equation}}
\newcommand{\bc}{\begin{center}}
\newcommand{\ec}{\end{center}}
\newcommand{\bdm}{\begin{displaymath}}
\newcommand{\edm}{\end{displaymath}}
\newcommand{\ds}{\displaystyle}
\newcommand{\p}{\partial}
\newcommand{\INT}{\int\limits}
\newcommand{\SUM}{\sum\limits}
\newcommand{\bfm}[1]{\mbox{\boldmath $ #1 $}}
\title{ \bf
Mathematical modelling of drug release \\ from multi-layer capsules
}

\author{
{\em  Badr Kaoui$^{a}$, Marco Lauricella$^{b}$, Giuseppe Pontrelli$^{b}$
\footnote{Corresponding author, E-mail: {\tt giuseppe.pontrelli@gmail.com}}}
\vspace{5mm}\\
$^{a}$  Biomechanics and Bioengineering Laboratory (UMR 7338)\\
 CNRS, Sorbonne University \\
University of Technology of Compi\`{e}gne, 60200 Compi\`{e}gne, France  
\vspace{1mm}\\
$^{b}$Istituto per le Applicazioni del Calcolo - CNR \\
Via dei Taurini 19 -- 00185 Rome, Italy \\
}

\maketitle


\abstract{
We propose a mathematical model for computing drug release from multi-layer capsules. 
The diffusion problem in such heterogeneous layer-by-layer composite medium is described by a system of coupled partial differential equations, which we solve analytically using separation of variables.
In addition to the conventional interlayer continuous mass transfer boundary conditions, we consider also the case of finite mass transfer resistance, which corresponds to the case of a coated capsule.
The drug concentration in the core and through all layers, as well as in the external release medium, is then given in terms of a Fourier series that we compute numerically to characterize the drug release mechanism.

\section{Introduction}
\setcounter{equation}{0}
Nowadays, there is a tremendous increasing interest in using capsules, beside other sub-micrometer bodies such as nanoparticles, as targeted drug delivery systems \cite{siebook}.
They allow enhancing therapeutic efficacy and reduce side effects by controlling the drug dose released in the human body.
Capsules consist usually of a drug-loaded core (fluid or solid) surrounded by more than one hydrogel layer.
Such encapsulation with multiple layers enhances the capsule mechanical stability, its biocompatibility, protects the active ingredients from external chemical aggression and premature degradation, and it allows extending the sustainability of the drug release \cite{kok,tim}. 
Different technologies have emerged in the last years to design and build layer-by-layer concentric spherical capsules \cite{has,lin}, even though the release of their encapsulated drug cannot be fully predetermined.  
For some specific applications, a thin coating shell (or membrane) is required to envelop the whole capsule structure in order to protect it from external chemical aggressions and mechanical erosion.
Depending on the nature of the encapsulated formulations and according to the final aimed therapeutic requirements the typical size of capsules can range from some nano-meters to milli-meters. 
For biomedical applications, micro-capsules are largely used \cite{mas}.  
 
Drug release characterization consists in tracking the kinetics of the drug eluted from the capsule into the external targeted medium, which is away from being an easy task in both cases \textit{in situ} and \textit{in vivo}.
However, effort and costs of developing and designing new optimized delivery devices can be dramatically reduced if the release mechanism is understood in advance using appropriate \textit{in silico} models \cite{lang,pep,pep1}. 
A number of review papers have summarized previously proposed models for drug release either from coated formulations \cite{gras,siep} or from polymeric matrices \cite{pep}. 
There are also reviews on drug release from capsules \cite{tim,hen,pha,tav}, most of them dealing with empirical models.
Here, we are rather interested in mechanistic models, because they allow a better understanding of the underlying mechanism of the drug release by tracking the influence of each physical input parameter, in contrast to the empirical models.
We upgrade existing mechanistic models by extending their applications to the multi-layer uncoated and coated capsules.
Thus, predictable simulations based on those models can spot the significant physical parameters and allow extracting reliable assessment for conducting \textit{in vitro}, \textit{in situ} and \textit{in vivo} experiments.
Diffusion is by far the dominant mechanism in drug delivery beside other physico-chemical factors, such as osmosis, drug dissolution, and polymer swelling \cite{gras}.  
Most of the previous suggested mathematical models for release from spheres, as well as dissolved
or dispersed drug systems, rely on oversimplifying assumptions, such as considering
a constant diffusion rate and a well stirred (or at constant concentration) release medium \cite{cra}, as summarized in the review by Arifin \textit{et al} \cite{ari}.  Mass flux resistance at the coating shell for the coated capsules is not properly addressed in literature.

In this article, we will make a step forward by studying theoretically for the first time the drug release from a multi-layer capsule, via a semi-analytic procedure, and moreover by avoiding considering simplistic hypothesis made elsewhere previously, such constant and continuous concentration at the capsule surface. 
A general presentation of a pure diffusive model through a composed layer-by-layer medium is given in Sect. 2. 
The specific case of a spherical core-layer capsule coated with a protective thin shell is addressed in Sect. 3, with a mathematical treatment
somehow similar to the one we previously used for other drug-eluting systems \cite{pon2,pon}.
Here we describe in detail this method and how to use it for the problem at hand.
Special care is taken when setting the interlayer conditions, including the finite resistance (jump in the concentration value) at the external coated shell.
An analytic expression for the concentration and the cumulative mass in all layers are given in Sect. 4. 
Finally numerical simulations are used to study the sensitivity of the system  to various parameters and configurations. 

\section{A general model for drug release from a multi-layer capsule}
Let us consider a multi-layer capsule made of a drug-filled core (the depot, $\Omega_0$) surrounded by a successive number of layers ($\Omega_i, \; \text{with} \; i=1,2,...,n$) as illustrated in Fig. 1. 
These enveloping layers are constituted of different, but homogeneous and isotropic materials, which are customized to allow selective diffusion, to better control the release rate and sometimes to host more than one drug \cite{mas}.
The last protective outer shell is in contact with the targeted release medium $\Omega_s$ (either a bulk fluid or tissue), here of finite extent, with boundaries "{\em far enough}" from the capsule surface. 
Even in such a simple configuration where the transport is driven by pure diffusion, predicting drug kinetics is not an easy task.

In the core $\Omega_0$, we assume that the drug dissolution occurs in a short time compared to that of diffusion and thus, can be considered as instantaneous \cite{gin}. 
Therefore, the drug diffusion in $\Omega_0$ can be described by the second Fick's law:
\be
{\p c_0 \over \p t} = D_0 \nabla^2 c_0   \, \,\mbox{in}\, \, \Omega_0,  \label{eq1}
\ee
where $c_0$ is the concentration field and $D_0$  is the diffusion coefficient of the drug in the core. 
Analogously, in the surrounding layers and in the release medium $\Omega_s$, we have similar diffusion equations 
\be
{\p c_i\over \p t} =D_i \nabla^2 c_i
\qquad \mbox{in}  \, \Omega_i  \qquad i=1,2,...,n,s   \label{eq31} 
\ee
with $D_i$ is the diffusion coefficient in the i-th layer $\Omega_i$ (any possible convection, as in Ref.~\cite{Kaoui2017} or reaction terms in $\Omega_s$  are excluded). 
We set a perfect sink condition far away (see Sect. 3):
\be 
\label{61}  c_s =0  \;\;   \, \mbox{at} \, \, \p \Omega_s, \quad t>0
\ee
The initial condition for concentrations:
\be
c_i(\cdot,0)=f_i( \cdot)     \qquad\qquad \mbox{in} \,\, \Omega_i  \qquad i=0,1,...,n,s 
\ee
are given in all layers.   \par

Here, we refrain from using two widely exploited over-simplistic assumptions \cite{cra,ari}, i.e.: \\
 i) $c_0$ is constant at core surface $\p \Omega_0$, as if there is a sustained source of drug;\\
ii) $c_s$ is uniform in the release medium $\Omega_s$, as in well-stirred medium.

\bigskip
\underline{Modelling interlayer boundary conditions} \\
At each interface between two adjacent layers, flux continuity holds:
\be
-D_i \nabla c_i  \cdot \bfm{n} =  -D_{i+1} \nabla c_{i+1} \cdot \bfm{n}  \qquad\qquad  \mbox{at}  \,\,  \p \Omega_i \cap \p \Omega_{i+1} \label{gh3} 
\ee
with $\bfm{n}$ the surface external normal vector.
Moreover, a non-perfect contact is present at the interlayers \cite{cus}:
\be
 c_i=\sigma_{i}c_{i+1}, \qquad\qquad  \mbox{at}  \,\,  \p \Omega_i \cap \p \Omega_{i+1}  \label{gh2}   
\ee
where $\sigma_{i}$ is the drug partition coefficient between layers $i$ and $i+1$.
\bigskip

\underline{Modelling the external coating shell}\\
To prevent fast delivery, the capsule's outmost layer $\Omega_n$  is protected with a thin semi-permeable shell ({\em coating}) $\Omega_m$  having a small, yet finite thickness ($h$), see Fig. 2. 
This coating shields and preserves the encapsulated drug from degradation and 
fluid convection,  protects the capsule structure, and guarantees
a more controlled and sustained release \cite{hen}. Instead of using a distributed model as above, we model the drug diffusion across $\Omega_m$ by setting a simple interface
boundary condition between  $\Omega_n$ and $\Omega_s$ that incorporates the physical properties of  the coating shell as follows. 
The mass flux across $\Omega_m$ is proportional to the concentration gradient:
\be
\bfm J_m= - D_m \nabla c_m   \label{ert}
\ee
where $D_m$ is the diffusion coefficient in the coating shell. 
Let $\sigma_n$ and  $\sigma_m$  be the partition coefficients (i.e. the ratio of the left/right concentrations at the equilibrium, see Fig. 2) of drug on both interfacial sides of the coating shell :
\be
\sigma_{n}={c_n \over c_{m}^-},  \qquad\qquad  \sigma_{m}={c_{m}^+ \over c_s}  \label{ert1}
\ee
By applying Eq. (\ref{ert}) across the coating shell thickness $h$ (fig 2, right zoom), we get:
\be
\bfm J_m \cdot \bfm n=-D_m {c_{m}^+ - c_{m}^-  \over h}= {D_m \over h}\left( {c_{n} \over \sigma_{n}}  - \sigma_{m} c_{s}  \right)=
{D_m \over h \sigma_{n}}\left( c_{n}  - \sigma_{n} \sigma_{m} c_{s}  \right) \label{ert2}
\ee
This can be viewed as an additional boundary condition, of different nature, imposed at the interface separating $\Omega_n$ and $\Omega_s$.  
Thus, in case of having coating shell, Eqns (\ref{gh3})--(\ref{gh2}) are then replaced with:
\be
-D_n \nabla c_n  \cdot \bfm n  = -D_s \nabla c_s \cdot \bfm n = P (c_n - \Sigma c_s)  \qquad\qquad \mbox{at} \,\,
 \p \Omega_n \cap \p \Omega_s   \qquad \label{eru34}
 \ee
where $P = \ds{D_m \over h \sigma_{n}}$ is mass transfer coefficient ($m s^{-1}$) and $\Sigma=  \sigma_{n} \sigma_{m}$.
 
Equation (\ref{eru34}) lumps the distributed model in $\Omega_m$  into a simpler interlayer boundary condition, where $P$ reflects the shell mass transfer properties into one single easily measurable coefficient that is related to the permeability \cite{cus}. 
At the coating shell surface, we use the interlayer finite resistance Eq. (\ref{eru34}) rather than Eq. (\ref{gh2}).

\section{A case study: a spherical core-shell capsule}
\setcounter{equation}{0}
The two-layer capsule is currently the most used and in this section we consider this special case: a drug-filled core  $\Omega_0$ 
encapsulated by a single polymeric shell $\Omega_1$ and surrounded by the release medium  $\Omega_2$: each medium has the shape of concentric sphere of increasing radius, $R_0, R_1, R_{\infty}$ respectively (with $ R_0< R_1 \ll  R_{\infty}$).
This method can be easily extended, with a more complicated algebra, to any number of layers. 
Due to the homogeneity and isotropy, we can assume that net drug diffusion occurs along the radial direction only, and thus we restrict our study to a one-dimensional model (Fig. 2) that reflects a perfectly radially symmetric system. 
Thus, the general formulation given in Section 2 reduces to a three coupled equations problem, that in $1D$ radial symmetry reads: 

\begin{align}
&{\p c_0 \over \p t} = D_0 \left( {\p^2 c_0 \over \p r^2} + {2 \over r} {\p c_0 \over \p r}\right)={D_0 \over r^2} {\p \over \p r}\left(r^2 {\p c_0 \over \p r}\right)
&\mbox{in}\, \, (0, R_0) \label{erf4}  \\
&{\p c_1 \over \p t} = D_1 \left( {\p^2 c_1 \over \p r^2} + {2 \over r} {\p c_1 \over \p r}\right) = {D_1 \over r^2} {\p \over \p r}\left(r^2 {\p c_1 \over \p r}\right)
&\mbox{in}\, \, (R_0, R_1)  \label{erf5} \\
&{\p c_2 \over \p t} = D_2 \left( {\p^2 c_2 \over \p r^2} + {2 \over r} {\p c_2 \over \p r}\right) ={D_2 \over r^2} {\p \over \p r}\left(r^2 {\p c_2 \over \p r}\right)
&\mbox{in}\, \, (R_1,  R_{\infty}) \label{erf6}  \\
&{\p c_0 \over \p r}=0   & \mbox{at} \, \, r=0  \label{erf7} \\
&-D_0{ \p c_0 \over \p r} = -D_{1}{ \p c_{1} \over \p r}, \qquad c_0= \sigma_{0} c_1   &\mbox{at}\, \, r=R_0 \label{erf8}   \\
 &-D_1{ \p c_1 \over \p r} = -D_{2}{ \p c_{2} \over \p r}, \qquad c_1= \sigma_{1}c_2  &\mbox{at} \, \, r=R_1    \label{erf9} \\
 &c_2 =0   &\mbox{at}\, \, r=R_{\infty}   \label{erf0}  
 \end{align}
where $r$ is the radial coordinate position. 
In case of having a coating shell $\Omega_m$ Eq. (\ref{erf9}) is replaced with (see Eq. (\ref{eru34})):
\be 
-D_1{ \p c_1 \over \p r} = -D_{2}{ \p c_{2} \over \p r}= P (c_1 - \Sigma c_2)   \qquad\qquad \mbox{at}\, \, r=R_1 \nonumber \\
\ee
The initial condition for a releasing capsule is:
\begin{align}
&c_0(r,0)=C_0    \qquad\qquad \mbox{for} \quad 0 < r \leq R_0 \nonumber \\
&c_1(r,0)=0     \qquad\qquad  \mbox{for} \quad   R_0 < r \leq R_1 \nonumber \\
&c_2(r,0)=0     \qquad\qquad \mbox{for}  \quad   R_1 < r \leq R_{\infty}  \label{bs1}
\end{align} 
with $C_0$ is the initial concentration of the loaded drug in the core.  
 
\bigskip
\underline{The release distance} \\
The release medium is here delimited by a cut-off finite distance $R_{\infty}$, whose range is determined by the following considerations. 
Strictly speaking, for a pure diffusion problem from a spherical source into a homogeneous medium having the diffusivity $D$, the concentration field $c(\cdot,t)$ undergoes a exponential decay $\propto \exp(-Dt)$.
On the other hand, at a given time, the concentration is gradually damped, going down to zero at infinite distance. 

For computational practical purposes, we set a cut-off length $R_{\infty}$, that we call {\em release distance } or {\em penetration depth}, beyond which the concentration and as well as the mass flux  
reduce to a given percentage of their initial values at a given time.
More precisely, the {\em release distance } is the minimum finite length such that:
$c(R_{\infty}, t) = \epsilon c_0 \simeq 0$ at time $t$, with $\epsilon$ a tolerance.  For $ 0 \leq r \leq R_{\infty}$ the concentration decays with time at exponential rate and  the sink condition (\ref{erf0}) holds at $ R_{\infty} $ within a given tolerance, being $c(r,t) \approx 0$ for $r \geq R_{\infty}$.  
Analogously to the heat transfer problem,  $R_{\infty}$ is proportional to $\sqrt{Dt}$ \cite{mil}. The precise estimation of the release distance  in a multi-layer medium is beyond the
scope of this work, even though there are attempts to compute it within a certain
degree of accuracy. 
However, by considering a homogeneous medium at constant concentration faced with the release medium, we get a conservative overestimation of the release distance by $R_{\infty} =\sqrt{10m D t}$, with $\epsilon=10^{-m}$ \cite{pon2}.

\bigskip 
\subsection{Solving procedure} 
We use the separation of variables method to solve the three-layer equation model, as done in other composite release systems \cite{pon2,pon}. 
To this purpose, it would be convenient to recast the above equations in a dimensionless form. 

\underline{\textit Scaling} \\
All the variables, the parameters and the equations are normalized by means of the change of variables:
\be
 r \rightarrow {r \over R_{\infty}}   
\qquad\qquad  t \rightarrow { D_{max} \over R_{\infty}^2} t 
\qquad\qquad  c_i  \rightarrow {c_i \over C_{0}} 
\ee
and by introducing the nondimensional constants:
\be
R_{0/1/\infty} =  { R_{0/1/\infty}  \over R_{\infty}} \qquad    
\qquad   \gamma_i={D_i \over  D_{max}}  \qquad\qquad  \phi= {P R_{\infty} \over D_{max} }     \label{gh6}
\ee
where subscript $\max$ denotes the maximum value across the $3$ 
layers. 
By separating $c_i(r,t)=F_i(r)G_i(t)$, Eqs. (\ref{erf4})--(\ref{erf6}) become:
\be
{1 \over \gamma_i} {G_i' \over G_i}=  {(r^2 F_i')' \over r^2 F_i} = -\lambda_i^2\qquad   i=0,1,2 \label{df3}
\ee
that admits the solutions:
\be
F_i(r)=a_i {\cos(\lambda_i r) \over r}+ b_i {\sin(\lambda_i r) \over r},  
 \qquad 
G_i(t)= \exp ( - \gamma_i \lambda_i^2 t) \qquad i=0,1,2 \label{cv1}
\ee
In order to retain a finite solution in $r=0$, we set $a_0=0$.
By imposing $G_0=G_1=G_2$, we obtain
\be
\lambda_i=\sqrt{\gamma_0 \over \gamma_i  } \lambda_0  \qquad i=1,2  \label{vb1}
\ee
The boundary condition  (\ref{erf7})  is automatically satisfied, whereas (\ref{erf8})--(\ref{erf0})  read respectively:
\begin{align}
&\gamma_0 F_0'(R_0)=\gamma_1 F_1'(R_0) \label{kl2} \\
&F_0(R_0)=\sigma_0 F_1(R_0)   \label{kl3} \\
&\gamma_1  F_1'(R_1) =\gamma_2 F_2'(R_1) \label{kl4} \\
& F_1(R_1)=\sigma_1 F_2(R_1)   \label{kl5}\\
&F_2 (1)=0  \label{kl6}
\end{align}
In case of having a coating shell $\Omega_m$, Eq.  (\ref{kl5}) has to be substituted in
\be
\gamma_1 F_1'(R_1) + \phi ( F_1 (R_1)- \Sigma F_2(R_1)) =0 \label{df1}
\ee
This set of $5$ algebraic equations (\ref{kl2})-(\ref{kl6}) form a homogeneous system with the unknowns $b_0, a_1, b_1, a_2, b_2$.
By imposing the coefficient matrix to be singular and by using Eq. (\ref{vb1}), we get a relationship in $\lambda_0$ (eigencondition). 
This problem is solved numerically with MATLAB by a successive bisection method \cite{dal}.  It admits an infinite number of roots $(\lambda_0^k)$ with $k=1,2,...$, and from them, the whole set of eigenvalues $(\lambda_i^k)$ with $i=1,2$ and $k=1,2,...$ are determined.
From each eigenvalue, the constants $a_i^k, b_i^k$ are obtained subsequently from Eqs. (\ref{kl2}) to (\ref{kl6}), and the eigenfunctions defined in Eq. (\ref{cv1}) have the form:   
\be
 F_i^k(r)=  a_i^k
{\cos(\lambda_i^k r) \over r} + b_i^k{ \sin(\lambda_i^k r) \over r}     \label{exp1}
\ee

\section{Computing drug concentration and mass}
\setcounter{equation}{0}
Once the eigenvalues $\lambda_0^k$  are computed, the corresponding time-variable functions $G_i^k$  defined by Eqs. (\ref{cv1}) are obtained as:
\be
G_0^k(t)= G_1^k(t)=  G_2^k(t)=\exp({- \gamma_0 (\lambda_0^k)^2  t})   \label{exp2}
\ee
Thus, the general solution of the problem (\ref{erf4}) - (\ref{bs1}) is given by a linear superposition of the fundamental solutions (\ref{exp1}) - (\ref{exp2}) in the form:
\be 
c_i(r,t)=\sum_{k=1}^{\infty} A_k F_{ik}(r) \: \exp({-\gamma_i(\lambda_i^k)^2  t})
\qquad i=0,1,2   \label{sl33}
\ee  
where the Fourier coefficients $A_k$ are computed in accordance with the initial conditions:
\begin{align}
&c_0(r,0)=1     \qquad\qquad \mbox{for} \quad 0 < r \leq R_0 \nonumber \\
&c_1(r,0)=0     \qquad\qquad  \mbox{for} \quad   R_0 < r \leq R_1 \nonumber \\
&c_2(r,0)=0     \qquad\qquad \mbox{for}  \quad   R_1 < r \leq 1  \label{cs1}
\end{align} 
By evaluating Eq.~(\ref{sl33}) at $t=0$ and multiplying it by $r^2 F_{im}$ we get after integration:
\be
\int\limits_{0}^{R_0} \sum_k A_k r^2  F_{0k}  F_{0m} \, dr =  \int\limits_{0}^{R_{0}} r^2 F_{0m}  \, dr     \qquad m=1,2,.... \label{ant3}
\ee
\be
\int\limits_{R_0}^{R_1} \sum_k A_k r^2  F_{1k} F_{1m} \, dr =   0      \qquad m=1,2,.... \label{ant4}
\ee
\be
\int\limits_{R_1}^{1} \sum_k A_k  r^2 F_{2k}  F_{2m}  \, dr = 0       \qquad m=1,2,.... \label{ant5}
\ee
By summing (\ref{ant3})  with  (\ref{ant4})  multiplied by $\sigma_0$ and with (\ref{ant5})  multiplied by $\sigma_0 \sigma_1$, and by the orthogonality of $F_{0k}, F_{1k}, F_{2k}$ (see appendix) we have:
\be
A_k \left( \int\limits_{0}^{R_0} r^2 F_{0k}^2 dr+ \sigma_0 \int\limits_{R_0}^{R_1} r^2 F_{1k}^2 dr+  \sigma_0  \sigma_1 \int\limits_{R_1}^{1} r^2 F_{2k}^2 dr \right) = \int\limits_{0}^{R_0} r^2 F_{0k}  dr 
\ee
By setting:
\be 
I_k=\int\limits_0 ^{R_0} \sin(\lambda_0^k r) r  dr= \left[{ \sin(\lambda_0^k r)
\over (\lambda_0^k)^2} - { r \cos (\lambda_0^k r) \over \lambda_0^k }  \right]_0^{R_0} = { \sin(\lambda_0^k R_0)
\over (\lambda_0^k)^2} - { R_0 \cos (\lambda_0^k R_0) \over \lambda_0^k }  
\ee
we have:
\be 
A_k=  {b_0^k I_k \over N_k}
\ee
with $N_k$ is the norm (see appendix).
The analytic form of the solution allows an easy computation of the dimensionless drug mass (per unit of volume) in each domain as a function of time as in Ref.~\cite{pon}:
\bdm
M_0(t)= \int\limits_{0}^{R_0} c_0(r,t) dr  \qquad\qquad 
M_1(t) =  \int\limits_{R_0}^{R_1}  c_1(r,t) dr
\qquad\qquad  M_2(t) =\int\limits_{R_1}^{1}  c_2(r,t) dr
\edm
By (\ref{sl33}) we have: 
\begin{align}
& M_0(t)= \sum\limits_{k} A_k  \;
 \exp(-\gamma_0(\lambda_0^k)^2  t) \int\limits_0^{R_0}  F_{k0}(r) dr \label{m0} 
\end{align}
Similarly:
\begin{align}
& M_1(t) = \sum\limits_{k} A_k \;
 \exp(-\gamma_1(\lambda_1^k)^2  t) \int\limits_{R_0}^{R_1}  F_{k1}(r) dr \label{m1} \\
& M_2(t) = \sum\limits_{k} A_k  \;
 \exp(-\gamma_2(\lambda_2^k)^2  t) \int\limits_{R_1}^{1}  F_{k2}(r) dr \label{m2}
\end{align}
In particular, we have: 
\be
M_0(0)=R_0 \qquad\qquad M_1(0)=M_2(0)=0
\ee
Moreover, Eqs. (\ref{m0}), (\ref{m1}) and (\ref{m2}) show that $\lim\limits_{t \rightarrow \infty} M_0(t) = M_1(t) =M_2(t)=0 $ .

\section{Results and discussion}
\setcounter{equation}{0}
The physical problem of drug release from a multi-layer capsule depends on a large number of parameters.
In this article, for simplicity, the following physical parameters are chosen to simulate the release of substance from an uncoated and a coated core-shell capsule.
These typical values of geometrical and diffusion parameters are taken from the work of Henning \textit{et al.} \cite{hen}:
\begin{align}
&R_0=1.5 \cdot 10^{-3} m  \qquad\qquad  R_1=1.7 \cdot 10^{-3} m     \qquad\qquad  \sigma_0=\sigma_1=1  \nonumber \\
& D_0=30 \cdot 10^{-11} m^2 s^{-1}  \qquad    D_1=5 \cdot 10^{-11} m^2 s^{-1}  \qquad  D_2=30 \cdot 10^{-11}  m^2 s^{-1}   \nonumber \\
& \mbox{and} \label{num1} \\ \nonumber
& P=5 \cdot 10^{-8}  m/s \qquad\qquad\quad  \Sigma=1   \qquad\quad \mbox{(for the coated capsule case)}  \nonumber
\end{align}
For these input numerical values, the estimated release distance $R_{\infty}=40 \cdot 10^{-3} m \approx 24 R_1$ is sufficient to have the complete drug depletion from the core and a negligible concentration by about $10$ days.
All the series appearing in the solution (see Eq.~(\ref{sl33}) and what follow) have been truncated at a finite number of terms $N=100$ for all times reported in the simulation.
A number of grid nodes proportional to the thickness of the layer to guarantee a convenient resolution has been considered.
For example, for the layer's sizes corresponding to the above parameters, a number of equidistributed $100$, $14$ and $2550$  points has been selected, respectively.

Drug is transported differently in each layer, that receives mass from the previous underneath layer and transmits it to the next above, in a cascade  sequence until being completely damped out at finite distance $r \leq 1$. 
Fig. 3 shows the concentration profiles in the case of uncoated and coated capsules, and reports lower values and a more uniform distribution in all layers in the first case.
Concentration is dropping down inside each layer, being possibly discontinuous at the interlayer interfaces, with the mass flux continuity preserved (Fig. 3). 
We also compute the fraction of drug mass retained in each layer,
defined as:
\be
\theta_i(t)={M_i(t) \over M_0(0)} \qquad\qquad  i=0,1,2
\ee
Table 1 gives different distribution of $\theta_i(t)$ in each layer, both in the case without and with the coating shell. 
Because we set the sink boundary condition (\ref{erf0}), a mass loss occurs out of the bound $r=1$ and it is transferred to the surrounding bulk medium ($M_e$)\footnote{
The remaining infinite release medium is denoted by the subscript {\em e}.}. In other words, due to the absorbing condition (\ref{erf0}), all drug mass is transferred at the external medium ($e$) at a sufficiently large time and the total mass is preserved and equals its initial value (say the drug mass in the coating $M_0(0)$), such that:
\be
{M_0(0)- \sum\limits_{i=0}^2 M_i(t) - M_e (t) \over M_0(0)} = 1 -  \sum\limits_{i=0}^2 
\theta_i(t) - \theta_e =0 
\ee  

\begin{table}[ht]
\caption{Percentage of the drug mass $\theta$ retained in each layer (including the external semi-infinite medium $e$) at different 
times. Left values are in the case without membrane, right values with membrane, red values indicate the peak mass. The reported differences concern mostly intermediate times, with the depletion time unaltered.}
\bc
{\footnotesize
\begin{tabular}{|| r | c | c | c | c ||}
\hline
 time ($\, s (\simeq d:h:m)$)  & $\theta_0 (\%) $ & $\theta_1  (\%) $ &$\theta_2  (\%) $ & $\theta_e  (\%)$  \\ \hline
$100 \; (\simeq 2m)$ & $96 - 96$   & $3.4 - 3.5$ & $0.07 - 0.006$  & $0.8 - 0.6$    \\ \hline
$500 \; (\simeq 8m)$ & $85-85  $ & ${\color{red}4.6} - 6.6$& $2.0 - 0.3$  & $8.8 -7.8$    \\ \hline
$1000 \; (\simeq 17m)$ & $72-76$ & $4.4-{\color{red}7.5}$  & $3.9-0.7$ & $20-16$   \\ \hline
$2000 \; (\simeq 33m)$ & $52-66$ & $3.6-7.5$ & $5.9-1.6$ & $38-25$  \\ \hline
$5 \cdot 10^3 \; (\simeq 1h:23m) $  & $25-53$ & $2.0-6.2$ & ${\color{red}7.2}-3.0$ & $66-37$   \\ \hline
$ 10^4 \; (\simeq 2h:47m) $ & $9.7-39$ & $0.9-4.5$ & $6.1-{\color{red}4.0}$ &  $83-53$ \\ \hline
$5 \cdot 10^4  \; (\simeq  13h:53m)$ & $0.6-3.8$ & $0.07-0.4$ & $1.9-2.4$  &  $97-93$ \\ \hline
$ 10^5 \; (\simeq 1d:4h) $ & $0.2-0.5$ & $0.03-0.06$ & $1.0-1.2$&  $98-98$  \\ \hline
$5 \cdot 10^5 \; (\simeq 5d:19h) $  & $0.02-0.02$ & $ 0.002 -0.002$ & $0.2-0.2$&  $99-99$  \\ \hline
$ 10^6 \; (\simeq 11d:14h) $ & $<10^{-2}-<10^{-2} $ & $<10^{-3} - <10^{-3}$ & $<10^{-1}- <10^{-1}$&  $>99->99$  \\ \hline
$5 \cdot 10^6 \; (\simeq 57d:21h) $ & $<10^{-6} - <10^{-6}$ & $ <10^{-6} -<10^{-6} $ & $<10^{-3} - <10^{-6}$&  $\approx 100 -\approx 100 $  \\ \hline
\end{tabular}
}
\ec
\end{table}

\bigskip

Due to the diffusive coefficient, the mass is exponentially decreasing in the core (layer 0), but is first increasing up to some upper bound and then decaying asymptotically in the layer 1 and 2  (Fig. 4). In the outmost layer ($e$) the mass accumulates over an extended distance as the time proceeds. 
The simulation points out the time of peak mass  in the observable release medium (layer 2) is $4700 s$ ($\approx$ 1h:18min) for the case without the coating shell, and $14800 s$ ($\approx$ 4h:7min) for the case with membrane, demonstrating a prolonged release in the latter case.
The thin layer $1$ retains a negligible mass due to its thickness, and the core  is completely emptied ($\theta(\%) < 10^{-2}$) after a period of about 11 days (Table 1). At that time, all the mass is transferred to the external layer. 
However, a much sustained release occurs in the case with membrane (Table 1 and Fig. 4) at intermediate times, with the final delivery time unchanged.

It appears that the relative size of the layers and their respective diffusivity affect the whole drug release processes. 
Thus, it is worth identifying which set of parameters guarantees a more prolonged and uniform release and what other values are 
responsible for a localized peaked distribution followed by a faster decay. 
One of them is the permeability of the coating shell that offers a significant resistance to the mass flux.
Thus, the membrane properties have to be properly tuned in order to allow drug molecules to be released, while maintained in the efficient therapeutic range with exceeding the toxic dose and without dropping drown below an insufficient dose.
These results can be used to assess whether drug targets tissues at the desired rate and to optimize the dose capacity given by thin
surface coating shells for an extended period of time.  
Differently than in other single layer models, the current formulation constitutes a simple tool to predict the accurate drug release from a multi-layer capsule (uncoated and coated ones) that can help in designing and in manufacturing new drug delivery platforms. 

\section{Conclusions}
Despite notable recent progress in designing and manufacturing drug release systems, the precise delivery characteristics and the 
development of multifunctional delivery carriers remains a challenge.
Mathematical modeling has emerged in recent years as an additional powerful alternative tool to simulate drug delivery processes and much effort is currently addressed for a deeper understanding of the elution mechanism.  
The release process from a capsule is not completely understood and can be influenced by different concurrent physical and chemical factors.
In this work we propose and use a mechanistic model for studying the drug release from a multi-layer capsule under a limited number of physical assumptions. 
The analytic form of the solution provides insights into the drug mass transfer as well as the effect of design parameters, such as the device geometry and drug loading, on the release mechanism. 
By showing the relationship among the several variables and material diffusive properties, the present model can be used to identify simple indexes or clinical indicators of biomedical significance and to optimize drug elution towards a desired targeted organ or tissue. 
Thus, the capsule design for a therapeutically optimal rate can be predicted using a systematic approach with a minimum number of experimental studies. 

\bigskip\bigskip 
\noindent \underline{\bf Acknowledgments} \\
We are grateful to E.J. Carr of the Queensland University of Technology, F. de Monte of the University of L'Aquila and R. Jellali of the University of Technology of Compi\`{e}gne for their valuable discussions and helpful comments. 
This work has been partially supported by a STSM Grant from EU COST Action 15120 {\em Open Multiscale Systems Medicine} (OpenMultiMed).

\section*{Appendix}
\renewcommand{\theequation}{A.\arabic{equation}}
\renewcommand{\thesubsection}{A.\arabic{subsection}}
\setcounter{equation}{0}
Let us prove the orthogonality of the system $F_i(r), \; i=0,1,2$ -- see Eq. (\ref{cv1}) -- in the interval $[0,1]$.
The Sturm-Liouville eigenvalue problems Eq. (\ref{df3}) can be written:
\begin{align}
& (r^2 F_0')'=-\lambda_0^2 r^2 F_0  \qquad \mbox{in} \,  [0, R_0]  \label{ml1} \\
&F_0'(0)=0   \label{gl1} \\
&\gamma_0 F_0'(R_0)=  \gamma_1 F_1'(R_0) \label{gl2} \\
& \nonumber \\
& (r^2 F_1')'=-\lambda_1^2 r^2 F_1  \qquad \mbox{in} \,  [R_0, R_1]  \label{ml2} \\
& F_0(R_0)=\sigma_0 F_1 (R_0)   \label{gl3} \\
&  \gamma_1 F_1'(R_1)=  \gamma_2 F_2' (R_1)\label{gl4} \\
& \nonumber \\
& (r^2 F_2')'=-\lambda_2^2 r^2 F_2  \qquad \mbox{in} \,  [R_1, 1] \label{ml3} \\
& F_1(R_1)=\sigma_1 F_2(R_1)  \label{gl5} \\
& F_2(1)=0  \label{gl6}
\end{align}
Let us consider two different eigenvalues $\lambda_{0m}$ and $\lambda_{0n}$ and the corresponding eigenfunctions $F_{0m}$, $F_{0n}$.
Multiplying Eq. (\ref{ml1}) by $F_{0n}$ and integrating:
\be
\lambda_{0m}^2 \int\limits_{0}^{R_0} r^2 F_{0m} F_{0n} dr= - \int\limits_{0}^{R_0}
 (r^2 F_{0m}')' F_{0n} dr= -\big[ r^2 F_{0m}' F_{0n} \big]_{0}^{R_0} + 
\int\limits_{0}^{R_0}
r^2 F_{0m}' F_{0n}' dr  \label{ort11}
\ee
Similarly, for the eigenvalue $\lambda_{0n}$
\be
\lambda_{0n}^2 \int\limits_{0}^{R_0} r^2 F_{0n} F_{0m} dr= - \int\limits_{0}^{R_0}
(r^2  F_{0n}')' F_{0m} dr= -\big[ r^2 F_{0n}' F_{0m} \big]_{0}^{R_0} + 
\int\limits_{0}^{R_0}
r^2  F_{0n}' F_{0m}' dr  \label{ort21}
\ee
Subtracting Eq. (\ref{ort21}) from  Eq. (\ref{ort11})  we have:
\be
(\lambda_{0m}^2-\lambda_{0n}^2) \int\limits_{0}^{R_0} r^2  F_{0n} F_{0m} dr=
-\big[r^2 F_{0m}' F_{0n} \big]_{0}^{R_0}+ \big[ r^2 F_{0n}' F_{0m} \big]_{0}^{R_0} \label{nm1}
\ee
Repeating a similar procedure for the Eq. (\ref{ml2}) (resp. Eq. (\ref{ml3})), for the  eigenvalues $\lambda_{1m},
\lambda_{1n} $ (resp. $\lambda_{2m},\lambda_{2 n} $ )  and  for the eigenfunctions $F_{1m},F_{1n}$ (resp. $F_{2m},F_{2n}$), we get:
\be
(\lambda_{1m}^2-\lambda_{1n}^2) \int\limits_{R_0}^{R_1} r^2 F_{1n} F_{1m} dr=
-\big[ r^2 F_{1m}' F_{1n} \big]_{R_0}^{R_1}+ \big[ r^2 F_{1n}' F_{1m} \big]_{R_0}^{R_1} \label{nm2}
\ee
\be
(\lambda_{2m}^2-\lambda_{2n}^2) \int\limits_{R_1}^{1} r^2 F_{2n} F_{2m} dr=
-\big[ r^2 F_{2m}' F_{2n} \big]_{R_1}^{1}+ \big[ r^2 F_{2n}' F_{2m} \big]_{R_1}^{1} \label{nm3}
\ee
Eq.  (\ref{nm1}) multiplied by $\gamma_0$  and by use of (\ref{gl1}) reads:
\be
\gamma_0 (\lambda_{0m}^2-\lambda_{0n}^2) \int\limits_{0}^{R_0} r^2 F_{0n} F_{0m} dr=
-\gamma_0 R_0^2 F_{0m}'(R_0) F_{0n}(R_0) + \gamma_0 R_0^2 F_{0n}'(R_0) F_{0m}(R_0) \label{re1}
\ee

Eq. (\ref{nm2})  multiplied by $\gamma_1 \sigma_0 $ gives:
\begin{align}
&\gamma_1 (\lambda_{1m}^2- \lambda_{1n}^2) \sigma_0 \int\limits_{R_0}^{R_1} r^2  F_{1n} F_{1m} dr= -\gamma_1 R_1^2 \sigma_0 F_{1m}'(R_1) F_{1n}(R_1) + \gamma_1 R_0^2 \sigma_0 F_{1m}'(R_0) F_{1n}(R_0)   + \nonumber  \\   
&   \gamma_1 R_1^2 \sigma_0 F_{1n}'(R_1) F_{1m}(R_1) - \gamma_1 R_0^2 \sigma_0 F_{1n}'(R_0) F_{1m}(R_0) \label{re2}
\end{align}

Finally  Eq. (\ref{nm3})  multiplied by 
$\gamma_2 \sigma_0 \sigma_1 $  and by use of (\ref{gl6})  becomes:
\be
\gamma_2 (\lambda_{2m}^2-\lambda_{2n}^2)\sigma_0  \sigma_1 \int\limits_{R_1}^{1} r^2 F_{2n} F_{2m} dr=  \gamma_2 R_1^2 \sigma_0 \sigma_1 F_{2m}'(R_1) F_{2n}(R_1)  - \gamma_2  R_1^2 \sigma_0 \sigma_1 F_{2n}'(R_1) F_{2m}(R_1) \label{re3}
\ee

By summing Eqs.(\ref{re1}), (\ref{re2}), (\ref{re3}) and by use of Eqs. (\ref{vb1}), (\ref{gl2}) and (\ref{gl4}) we get:

\begin{align}
&\gamma_0 (\lambda_{0m}^2- \lambda_{0n}^2) \left( \int\limits_{0}^{R_0} r^2 F_{0n} F_{0m} dr+ \sigma_0 \int\limits_{R_0}^{R_1} r^2 F_{1n} F_{1m} dr+  \sigma_0  \sigma_1 \int\limits_{R_1}^{1} r^2 F_{2n} F_{2m} dr \right) =\nonumber \\
& \gamma_0  R_0^2  F_{0n}' (R_0)[F_{0m}(R_0)  -\sigma_0 F_{1m}(R_0)] - \gamma_0  R_0^2  F_{0m}' (R_0)[F_{0n}(R_0)  -\sigma_0 F_{1n}(R_0)] \nonumber \\
& + \gamma_1 R_1^2 \sigma_0 F_{1n}' (R_1)[F_{1m}(R_1)  -\sigma_1 F_{2m}(R_1)] - \gamma_1 R_1^2 \sigma_0 F_{1m}' (R_1)[F_{1n}(R_1)  -\sigma_1 F_{2n}(R_1)] \nonumber \\
& \label{jk1}
\end{align}
Finally, by the use of  (\ref{gl3})  and  (\ref{gl5}), all terms
at r.h.s. are zero and

\be
\int\limits_{0}^{R_0} r^2 F_{0n} F_{0m} dr+ \sigma_0 \int\limits_{R_0}^{R_1} r^2 F_{1n} F_{1m} dr+  \sigma_0  \sigma_1 \int\limits_{R_1}^{1} r^2 F_{2n} F_{2m} dr   = \left\{
\begin{array}{l}
0  \qquad\quad\, \mbox{for} \quad m \neq n  \\
N_m \qquad  \mbox{for} \quad m = n
\end{array}
\right.  \label{ort100}
\ee
with 
\begin{align}
N_m=&\int\limits_{0}^{R_0} (b_0^m \sin (\lambda_0^m r))^2 dr + \sigma_0 \int\limits_{R_0}^{R_1} 
 (a_1^m \cos(\lambda_1^m r )+b_1^m \sin (\lambda_1^m r))^2 dr \nonumber  \\
 + &\sigma_0  \sigma_1 \int\limits_{R_1}^{1}  (a_2^m \cos(\lambda_2^m r )+
b_2^m \sin (\lambda_2^m r))^2 dr  >0
\end{align}
 the norm.

\newpage
\begin{figure}[ht!]
\centering\scalebox{0.4}{\includegraphics{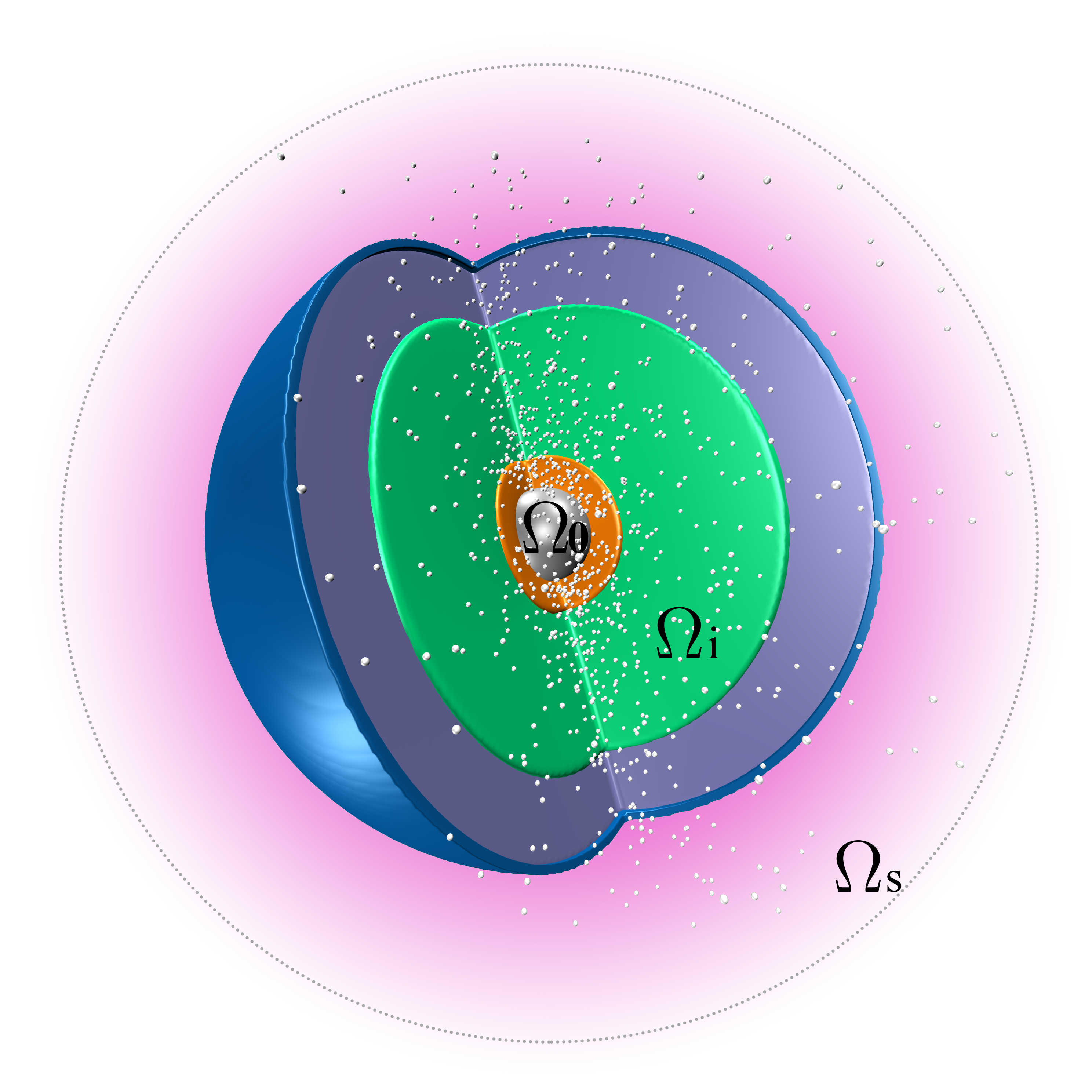}}
\caption{Drug releasing from a multi-layer microcapsule. Drug is
initially loaded in the core $\Omega_0$ and diffuses, through all the 
intermediate layers $\Omega_i $,
into the release  medium $\Omega_s$ confined by the dashed line.
Suitable conditions are set at internal interlayer interfaces and at the external
coating shell (figure not to scale).}  
\end{figure}

\bigskip \bigskip \bigskip 
\begin{figure}[ht!]
\centering\scalebox{0.6}{\includegraphics{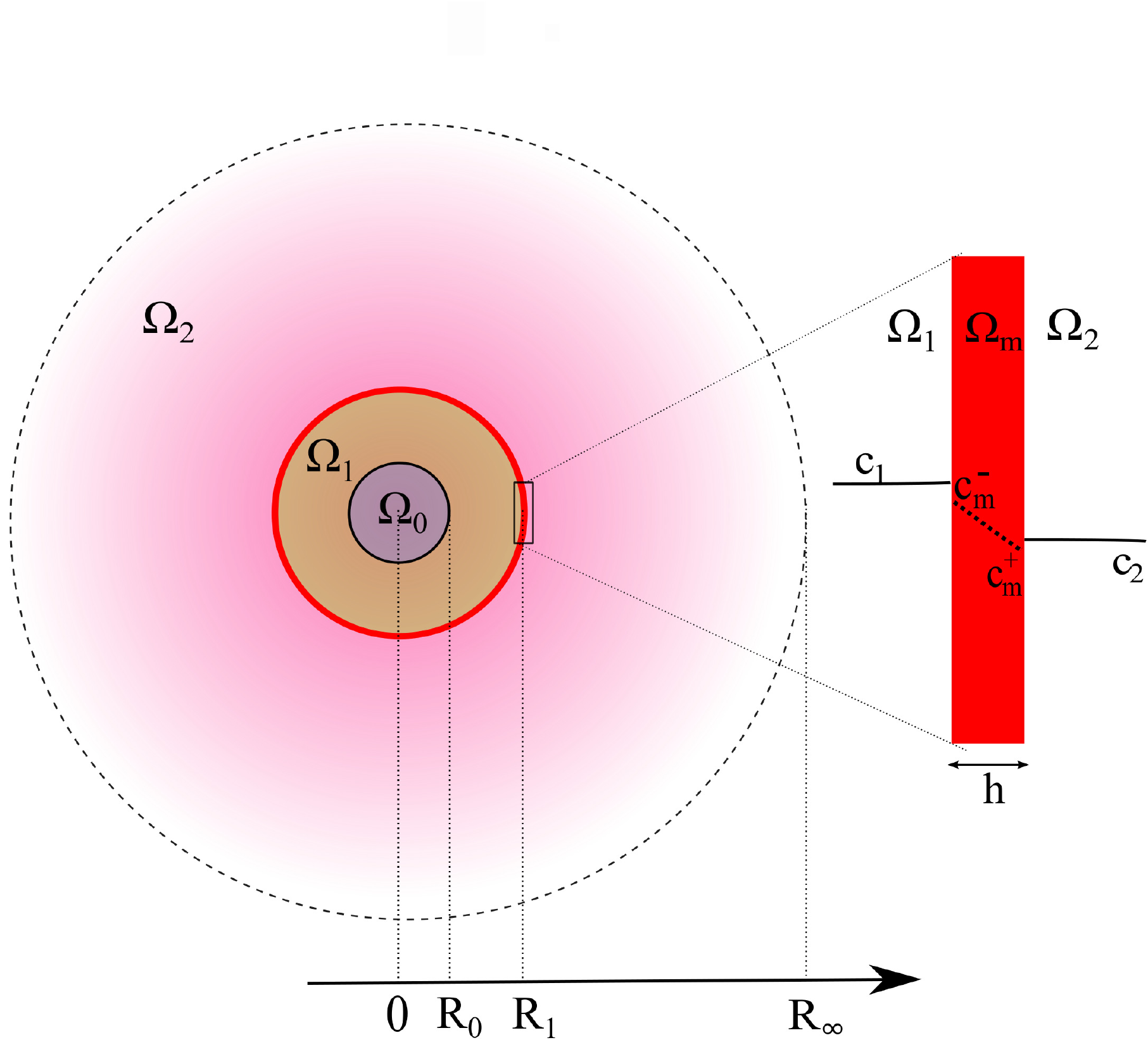}}
\caption{Schematic representation of the cross-section 
of the radially symmetric
multi-layer capsule, made of a central core $\Omega_0$, the concentric layer $\Omega_1$ and the possible thin coating shell $\Omega_m$ (in red). 
Together with  the external release medium  $\Omega_2$ - limited by the dashed line - it constitutes
a three concentric layers system. 
On the right side, a zoom  of the distributed coating layer shell (figure not to scale).}  
\end{figure}

\begin{figure}[ht!]
\centering\scalebox{0.7}{\includegraphics{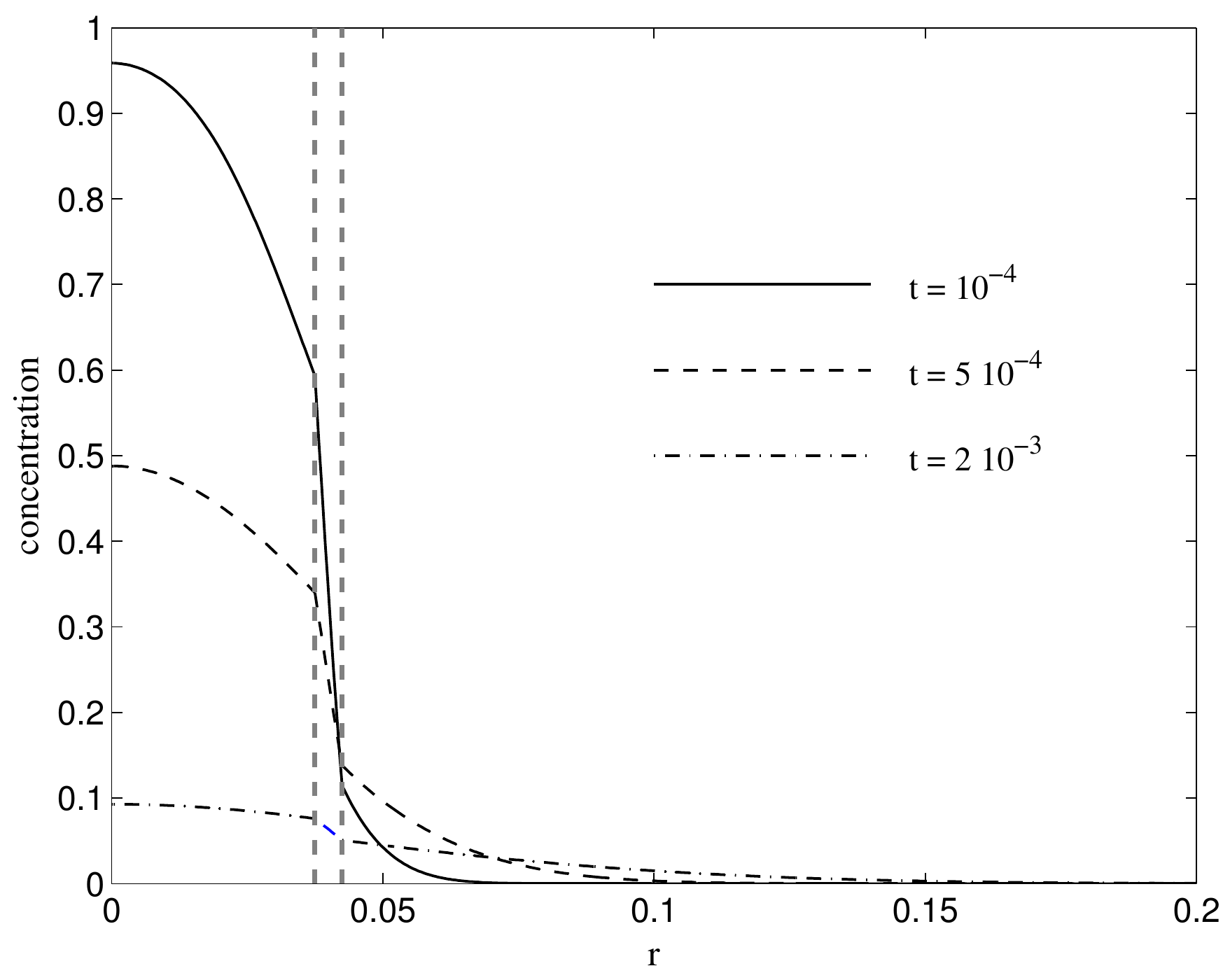}}   \\
\centering\scalebox{0.7}{\includegraphics{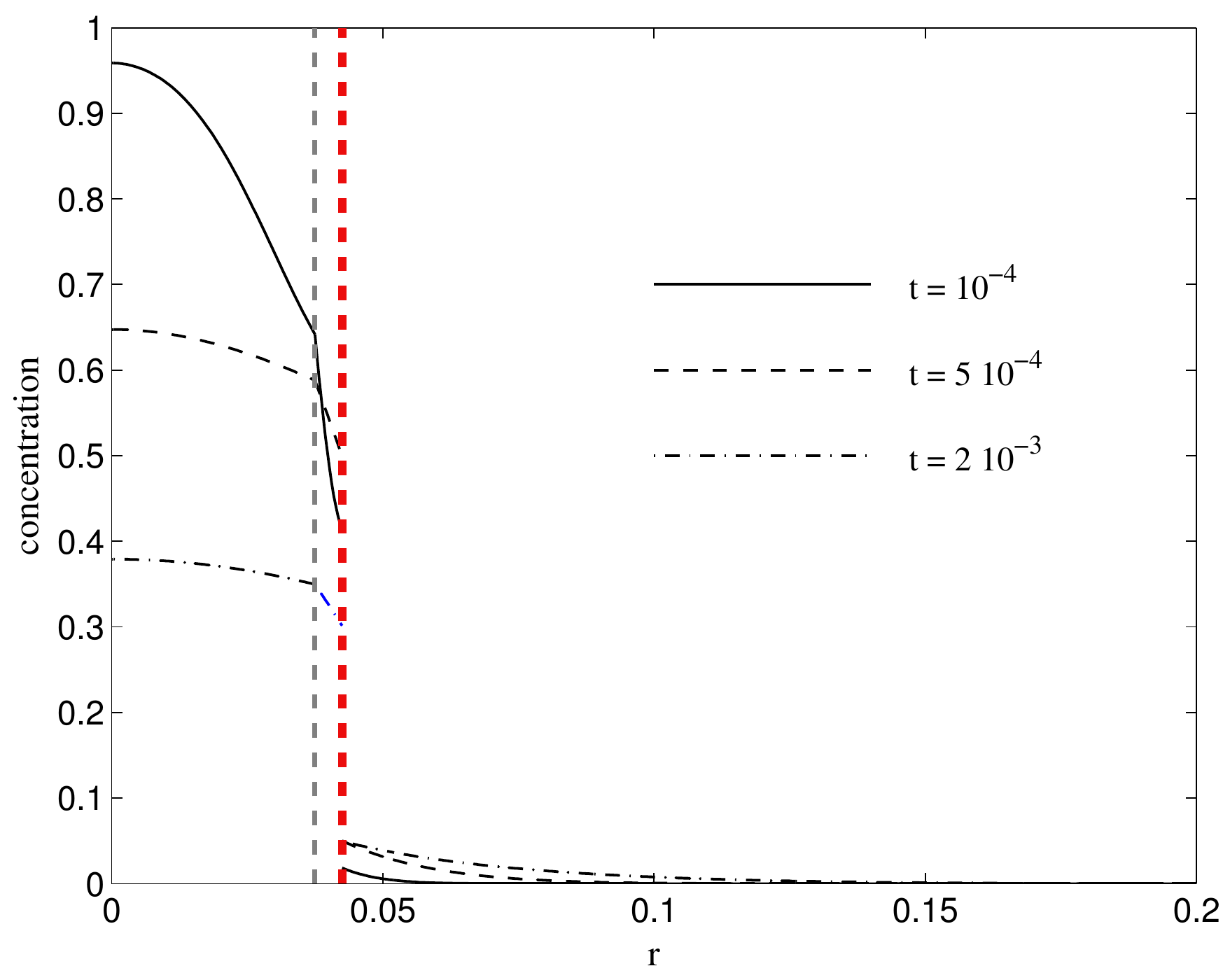}}
\caption{Normalized concentration profiles  in the three
  layers  at three instants, in case without (top) and with coating shell (bottom). At all observable times, all concentrations are
damped
out at nondimensional distance $r=0.2$. Note the slower release for the case with the coating shell.  }  
\end{figure}

\begin{figure}[h!]
\centering\scalebox{0.7}{\includegraphics{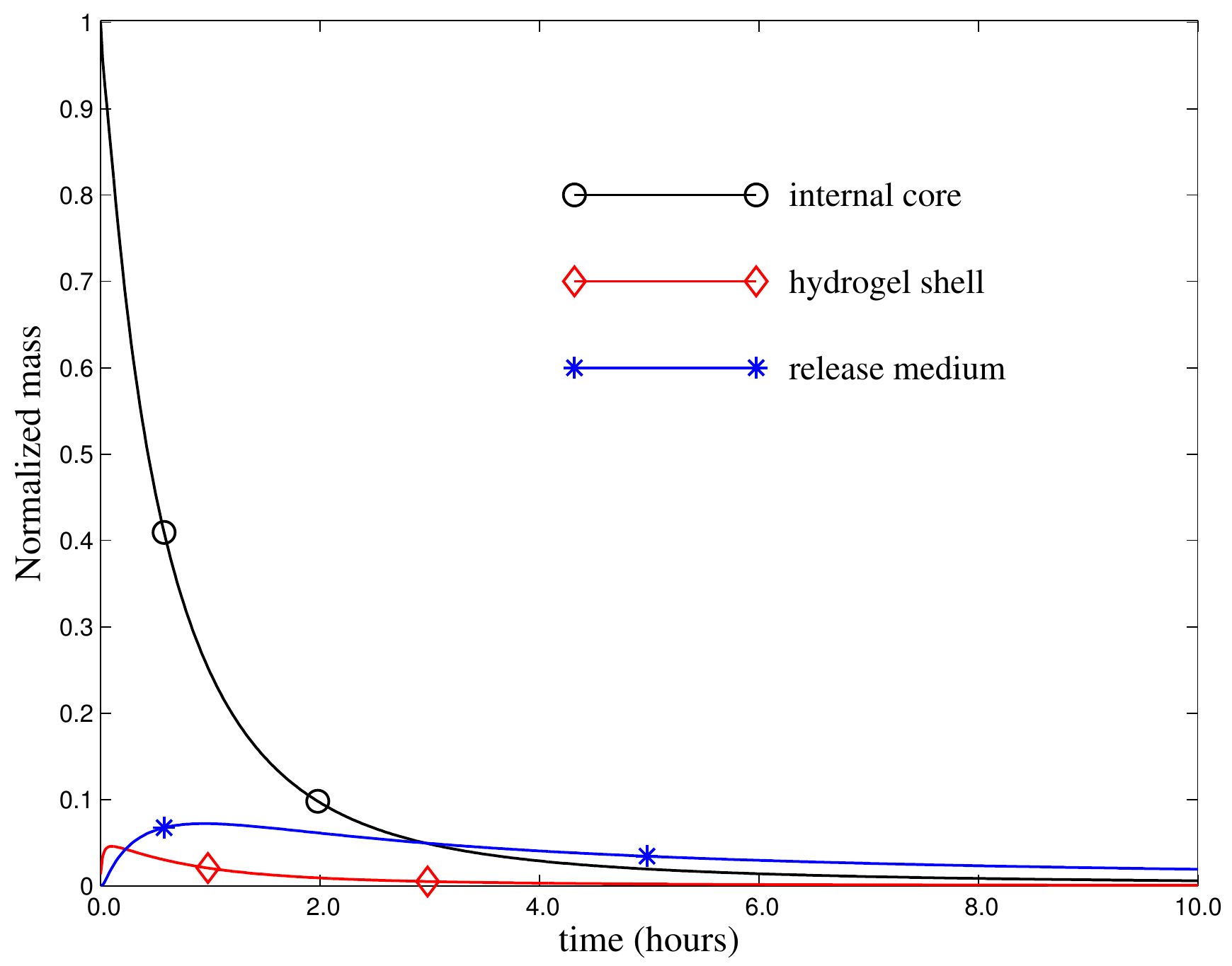}}  \\
\centering\scalebox{0.7}{\includegraphics{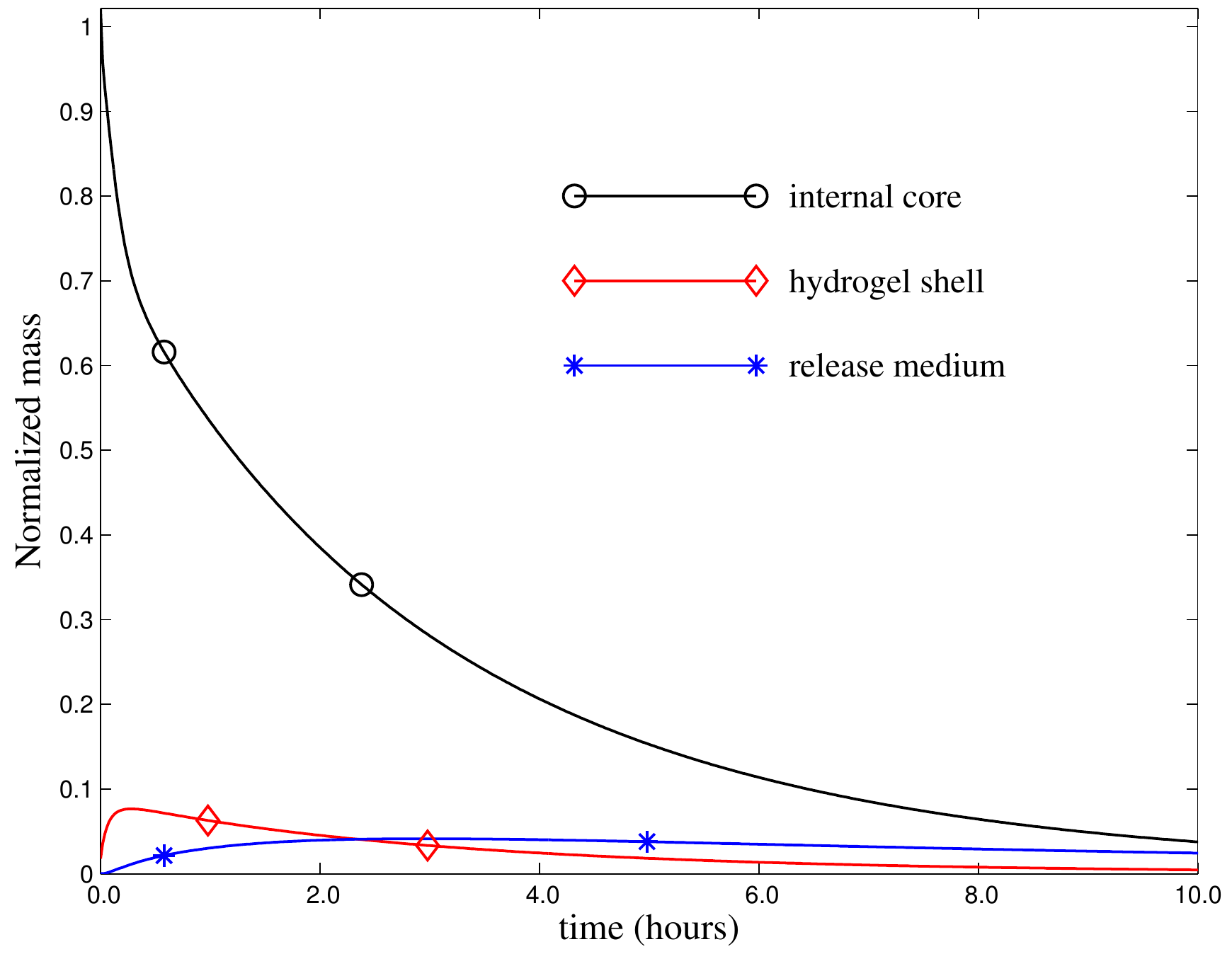}}  \\
\caption{Drug mass in the internal core (layer 0), in the layer (layer 1) and in the targeted release
medium (layer 2) in both cases without the coating shell (top), and with (bottom). In the internal layer mass is monotonically decreasing, while in the others
two there is a characteristic time 
at which the drug reaches a peak. Note a more uniform and sustained release 
in the second case.}  
\end{figure}

\end{document}